\documentstyle{article}

\if@twoside
\else
\fi
\advance\textheight by \topskip

\makeatletter
\def\@oddhead{\hfill\verb+aipproc2.tex+}
\let\@evenhead\@oddhead
\def\se{\vskip3pt plus1pt minus1pt\setbox0=\hbox to\hsize\bgroup\hss
        \vrule width.5pt
        \vbox\bgroup \hrule width \hsize height.5pt
        \vskip3pt\hbox to\hsize\bgroup\hss\vbox\bgroup\advance\hsize by-9pt
        \columnwidth\hsize\small}
\def\ee{\par\egroup\hss\egroup\vskip3pt\hrule width\hsize height.5pt\egroup
        \vrule width.5pt\hss\egroup
        \box0 \vskip3pt plus1pt minus1pt}
\def\latexe{\LaTeX\kern.15em 2${}_{\textstyle\varepsilon}$}

\makeatother

\begin{document}
\sloppy		

\begin{titlepage}
\vspace*{0pt plus1fill}
\begin{center}
\LARGE\bf
Leptoquarks\\[1pc]
Paul H. Frampton
\end{center}
\vspace*{0pt plus1fill}
\vspace*{0pt plus1fill}
\hbox to\hsize{\hfil
\vbox{\offinterlineskip\hrule height 2pt\vskip4pt
\Large\sf\setbox0=\hbox{May 4, 1997. Balestrand, Norway}\hbox to\wd0{\hfil Talk at Beyond the Standard Model V,\hfil}
\vskip4pt
\box0
\vskip4pt \hrule height 2pt}%
\hfil}
\end{titlepage}

\newpage

\begin{center}
\LARGE\bf
Abstract
\vspace{2pc}
\end{center}

In this review there is first a survey of the HERA data. Then we discuss
the theory response including querying the consistency of the data,
compositeness, contact terms, and leptoquarks including as a special case the R symmetry
violating squark. The SU(15) possibility for a light leptoquark is mentioned.
Finally there is a summary.

\newpage

\def\contentsname{\LARGE\bf Contents}

{\large
\tableofcontents
}

\twocolumn

\section{Data from HERA}

In the second half of February 1997, the two collaborations H1\cite{H1}
and ZEUS\cite{ZEUS} working on $e^+p$ collisions at HERA: $e^+$(27GeV) +
p(820GeV) simultaneously submitted to Z.Phys. announcements
of small-statistics discrepancies from the Standard Model(SM).
The two papers can also be downloaded from the World Wide 
Web at the URL: http://info.desy.de/

From the Web we may learn much interesting information. 
For example H1 has 400 members from 12 countries while ZEUS
has 430 members from 12 countries.

As for physics, H1 finds 12 events with $Q^2 > 15,000$GeV$^2$
where the SM predicts $4.71\pm0.76$. Coincidentally, ZEUS finds
2 events with $Q^2 > 35,000$GeV$^2$ for which the SM prediction is
$0.145\pm0.013$. The probability that these data result from a 
statistical fluctuation is 0.5\%. This is the same as the probability
that four dice thrown together will roll to a common number.

Another provocative fact, although not as clearcut when H1 and
ZEUS are compared, is that these excess high-$Q^2$ events {\it may}
cluster around a common {\it x} value. The mass $M = \sqrt{xs}$
of a direct-channel resonance would be $\sim200$GeV for H1 and 
perhaps somewhat higher for ZEUS. 

There are 20 times more data in $e^+p$ collisions compared to
$e^-p$. For the latter there has been an ion trapping problem at HERA.
The integrated luminosities are $34.3pb^{-1}$ of $e^+p$ data
and $1.53pb^{-1}$ of $e^-p$ data.

The conservative reaction is that (i) we have been chasing
deviations from the SM for two decades and experience has shown that
more precise data always remove the discrepancy. (ii) instead
of extrapolating the QCD to higher $Q^2$ we should make an overall fit
including the new events. Actually (ii) seems unlikely to work
because
here we are dealing with valence quarks, not gluons, with 
structure functions that are better known - it is not like the
situation for high transverse energy jets at FermiLab where
the gluon structure functions could be modified to explain the data.

For the present purposes, we take the experimental result seriously
at face value.

\section{Theory Response}

There have been 30 theory papers involving 63 theorists
in the 10 weeks
since the HERA announcement which for 14 events corresponds
to over 2 papers and 4 theorists per event! 

The 30 papers, with author(s) and hep-ph/97mmnnn numbers, 
are listed in chronological order of appearance in Refs.[3-32].


These 30 papers break down as follows:

\begin{itemize}

\item{ }1 on consistency of the data.\cite{dree}
\item{ }2 on compositeness.\cite{adl,aka}
\item{ }5 on contact interactions.\cite{bar,gon,bart,nel,buc}
\item{ }9 on R-symmetry breaking squark.
\cite{cho,alt,dre,kal,cho2,kal2,dre2,kon,barb}
\item{ }13 on light leptoquark.
\cite{kuo,fio,don,blu,bab,suz,hew,leo,pap,kun,ple,fri,mon}\\

\end{itemize}

\noindent
We see that a  new light ($\sim200$GeV) particle is the most popular
explanation with 22 papers out of the 30.\\

First, I shall review these 30 papers written by theorists in 1997.\\

Then I will review my own 1992 paper \cite{frampton}
predicting weak-scale leptoquarks.

\subsection{Consistency of Data}

This is from Drees, hep-ph/9703332. He compares H1 with ZEUS. First look at the
distribution of $M(LQ) = \sqrt{xs}$. One finds $200.3\pm1.2$ GeV (7 events)
for H1 and $219.3\pm5.5$ GeV (4 events) for ZEUS. These disagree by
$3.4\sigma$ but overall uncertainty in the energy reduces this to $1.8\sigma$,
or an 8\% chance occurrence.

Next compare the absolute event numbers 7 for H1 and 4 for ZEUS. The integrated
luminosities are $14.2pb^{-1}$ for H1 and $20.1pb^{-1}$ for ZEUS. Zeus has "looser" cuts
and the likelihood of this outcome is not higher than 5\%.

Combining the two effects (x and \# events) gives a 0.5\% compatibility between
H1 and ZEUS. This is about the same as the compatibility of the combined HERA data with
the Standard Model. So the size of statistical fluctuation is comparable!  
 
\subsection{Compositeness}

Adler (hep-ph/9702378) is the first 1997 theory of the HERA effect. In
his "frustrated SU(4)" for preons, the positron interacts with a gluon and 
makes a transition to a $E^+$ state, a kind of leptogluon which decays into a $e^+$
and a jet.\\

Akama, Katsuura and Terazawa (hep-ph/9704327) revive a 1977 model
and calculate cross-sections for several different composite states. The
conclusion is that certain composites are possible:
\begin{itemize}
\item{ }leptoquark, $e^{+*}$
\end{itemize}
while excluded are:
\begin{itemize}
\item{ }$Z^*$,$q^*$ (by pre-existing mass bounds)
\end{itemize}

\subsection{Contact Interactions}

A general approach is due to Buchmuller and Wyler (hep-ph/9704317)
who parametrize physics beyond the SM by four-fermion 
contact interactions; quark-lepton universality, absence 
of FCNC color-independence, flavor-independence and consistency with
atomic parity violation results are imposed. There is then a unique
current-current form of contact interaction:
\begin{equation}
\frac{\epsilon}{\Lambda^2} J^{\mu}J_{\mu}
\end{equation}
where:
\begin{equation}
J^{\mu} = \bar{u}\gamma^5\gamma^{\mu}u + \bar{d}\gamma^5\gamma^{\mu}d + 
\bar{e}\gamma^5\gamma^{\mu}e + \bar{\nu}\gamma^5\gamma^{\mu}\nu 
\end{equation}
This contact term shares the U(45) invariance of the standard model
when the gauge and Yukawa couplings vanish.

Rare processes like $K_L \rightarrow e^-\mu^+$ and LEP2 data constrain
$\Lambda > 5$TeV. The HERA data require, on the other hand, 
$\Lambda < 3$TeV. So Buchmuller and Wyler conclude
that {\it if} there is new physics there must 
be sizeable breaking of quark-lepton symmetry and/or new particles.\\

Nelson(hep-ph/9703379) looks at the atomic parity violation constraints.\\

Di Bartolomeno and Fabbrichesi(hep-ph/9703375) {\it and}
Barger, Cheung, Hagiwara and Zeppenfeld (hep-ph/9703311)
concur that $\Lambda < 3$TeV is needed to fit the HERA data.\\

Gonzales-Garcia and Novaes (hep-ph/9703346) examine 
constraints on contact interactions from the one-loop
contribution to $Z\rightarrow e^+e^-$.

\subsection{Leptoquark [R Breaking Squark]}

As mentioned above, 22 of the 30 papers in 1997 are in this direction.
From here on we assume a direct-channel resonance in $e^+q$.\\

\noindent
Note that a {\it valence} quark is more likely than a {\it sea} antiquark
because the latter is going to give $e^-p$ effects conflicting with
the data, despite the twenty times smaller integrated luminosity.\\

Also: a scalar is more likely than a vector. A
vector leptoquark has a coupling $\phi_{\mu}^{\dagger}G^{\mu\nu}\phi_{\nu}$ 
to gluons and would be produced by $\bar{q}q 
\rightarrow g \rightarrow \phi^{\dagger}\phi$ at the Tevatron.
We expect therefore the leptoquark to be a scalar corresponding
to $(e^+u)^{5/3}$ and/or $(e^+d)^{2/3}$.

There are seven scalar leptoquark couplings to 
$e^{\pm}q$, three of which involve $e^+q$. 
All demand that the LQ is an SU(2) doublet:
\begin{equation}
\mathcal{O} = \lambda_{ij}L_i\Phi d_j^C \\
\end{equation}
\begin{equation}
\mathcal{O}^{'} = \lambda_{ij}^{'}L_i \Phi^{'} u_j^C \\
\end{equation}
\begin{equation}
\mathcal{O}^{''} = \lambda_{ij}^{''}Q_i e^C_j \Phi^{''}
\end{equation}
For $\mathcal{O}$ to explain HERA data,
$\lambda_{11}\sim0.05$.

The branching ratio $B(K_L\rightarrow e^+e^-\pi^0)
\leq4.3\times10^{-9}$ implies that
$\lambda_{11}\lambda_{12}\leq1.5\times10^{-4}$.
This means $\lambda_{12}\leq3\times10^{-3}$, an unusual flavor hierarchy.\\

The rare decays $K^+\rightarrow \pi^+\nu\bar{\nu}$ and 
$\mu\rightarrow e\gamma$ also constrain the off-diagonal
$\lambda_{ij}$.

The conclusion about the flavor couplings of the leptoquark is:

\begin{itemize}
\item{} The leptoquark scalar needs to couple nearly diagonally
to mass-eigenstate quarks.
\end{itemize}

If one believes in weak-scale SUSY, a special case of scalar leptoquark
is an R-symmetry breaking squark. The quantity $R=(-1)^{3B+L+2S}$
clearly must be broken because $e^+q$ has $R=+1$ and $\tilde{q}$ has $R=-1$.\\

In the superpotential $\lambda_{ijk} L_i L_j \bar{E}_k + 
\lambda^{'}_{ijk} L_iQ_kD_k + \lambda^{''}_{ijk} U_iD_jD_k$,
we put $\lambda^{''}_{ijk}=0$ for B conservation.
Then to explain the data we need $\lambda^{'}_{1j1} \sim 0.04/\sqrt{B}$ 
where $B(\tilde{q} \rightarrow e^+d)$ is the branching ratio.\\

Neutrinoless double beta decay $(\beta\beta)_{0\nu}$ requires 
$\lambda^{'}_{111} < 7\times10^{-3}$ which excludes $\tilde{q}=\tilde{u}$.\\

Can the squark be $\tilde{q}=\tilde{c}$? If so, it implies that the rare decay,
$K^+ \rightarrow \pi^+\nu\bar{\nu}$, being measured at Brookhaven
is close to its current bound, which is interesting.\\

Finally, if $\tilde{q}=\tilde{t}$, 
atomic parity violation implies 
that $|\lambda^{'}_{131}| < 0.5$, so the branching ratio B can 
be much less than 1 and still be consistent. 

\section{The SU(15) Possibility}

The paper\cite{frampton}, published in 1992,
{\it predicts} light leptoquarks in $e^-p$, the mode
in which  HERA was then running.

Such scalar leptoquarks predicted by SU(15)
lie at the {\it weak} scale.
The situation can be contrasted with SU(5) grand unification where
"leptoquark" gauge bosons couple to ${\bf \bar{5}}$ and {\bf 10},
are simultaneously "biquarks", and hence must have mass $\sim 10^{16}$GeV.\\

At the Warsaw Rochester Conference in July 1996, the
speakers from HERA and Tevatron report failure to find leptoquarks
so I thought my idea might be wrong.\\

Now it is worth spending a few minutes to
review SU(15).

The SU(15) GUT was inspired by the desire to {\it remove proton decay}.
With:
\begin{equation}
15 = (u_1, u_2, u_3, d_1, d_2, d_3, \bar{u}_1, \bar{u}_2,
\bar{u}_3, \bar{d}_1, \bar{d}_2, \bar{d}_3, e^+, \nu, e^-)
\end{equation}
three times, every gauge boson has definite B and L which are therefore
conserved in the gauge sector.

The symmetry breaking may be assumed to follow the steps:
\begin{equation}
SU(15) \stackrel{M_{GUT}}{\rightarrow} SU(12)_q \times SU(3)_l
\end{equation}
\begin{equation}
\stackrel{M_B}{\rightarrow} SU(6)_L \times SU(6)_R \times U(1)
\end{equation}
\begin{equation}
\stackrel{M_A}{\rightarrow} SU(3) \times SU(2) \times U(1)
\end{equation}
$M_{GUT}$ can be as low as $6 \times 10^6$GeV, for $M_A=M_W$.
Once we select $M_A$, the other two scales $M_B$ and $M_{GUT}$ 
are calculable from the renormalization group equations.

In SU(15), the family-diagonal scalar
leptoquark is in the {\bf 120} representation. The content of
{\bf 120} is shown in \cite{frampton}. It contains $(3, 2)_{Y=7/6}$
under $SU(3) \times SU(2) \times U(1)$.
This SU(2) doublet has the quantum numbers of $(e^+u)$ and $(e^+d)$,
required to explain the HERA data.

The point is that this irreducible representaion contains the standard 
Higgs doublet so there is every reason to expect the scalar leptoquarks
to lie near the weak scale $\sim 250$GeV. SU(15) predicts further
leptoquark and "bifermion" states at or near the weak scale.\\

The discovery of light leptoquarks would be evidence as compelling
as proton decay for grand unification.

\section{Summary}

The HERA data (H1 and ZEUS Collaborations) suggest an s-channel
scalar resonance at $\sim200$GeV in $e^+q$.

There are 30 papers to explain 14 events.\\

The consistency of the H1 and ZEUS data has been questioned.

Possible explanations from 1997 include:

\begin{itemize}

\item{ } Compositeness
\item{ } Contact interactions.
\item{ } Leptoquark (R breaking squark).\\

\end{itemize}

In 1992:
\begin{itemize}
\item{ }An SU(15) GUT predicted light scalar leptoquarks.
\end{itemize}

\section*{Acknowledgement}

This work was supported in part by the US Department of Energy
under Grant No. DE-FG02-97ER41036.

\end{document}